\documentclass[prb,preprint,longbibliography]{revtex4-1} 

\usepackage{amsmath} 
\usepackage{longtable} 
\usepackage{graphicx} 
\usepackage{epstopdf} 
\usepackage{gensymb} 
\usepackage{textcomp} 

\begin{document}

\title{Model-Based Reasoning in the Upper-Division Physics Laboratory: Framework and Initial Results}

\author{Benjamin M. Zwickl}
\affiliation{School of Physics and Astronomy, Rochester Institute of Technology, Rochester, NY 14623}
\email{benjamin.m.zwickl@rit.edu} 
\author{Dehui Hu}
\affiliation{School of Physics and Astronomy, Rochester Institute of Technology, Rochester, NY 14623}
\author{Noah Finkelstein}
\affiliation{Department of Physics, University of Colorado Boulder, Boulder, CO 80309}
\author{H. J. Lewandowski}
\altaffiliation[Also at ]{JILA, University of Colorado Boulder, Boulder, CO 80309} 
\affiliation{Department of Physics, University of Colorado Boulder, Boulder, CO 80309}
\date{\today}

\begin{abstract}
We review and extend existing frameworks on modeling to develop a new framework that describes model-based reasoning in upper-division physics labs.  Constructing and using models are core scientific practices that have gained significant attention within K-12 and higher education. Although modeling is a broadly applicable process, within physics education, it has been preferentially applied to the iterative development of broadly applicable principles (e.g., Newton's laws of motion in introductory mechanics).  A significant feature of the new framework is that measurement tools (in addition to the physical system being studied) are subjected to the process of modeling.  Think-aloud interviews were used to refine the framework and demonstrate its utility by documenting examples of model-based reasoning in the laboratory.  When applied to the think-aloud interviews, the framework captures and differentiates students' model-based reasoning and helps identify areas of future research.  The interviews showed how students productively applied similar facets of modeling to the physical system and measurement tools: construction, prediction, interpretation of data, identification of model limitations, and revision. Finally, we document students' challenges in explicitly articulating assumptions when constructing models of experimental systems and further challenges in model construction due to students' insufficient prior conceptual understanding. A modeling perspective reframes many of the seemingly arbitrary technical details of measurement tools and apparatus as an opportunity for authentic and engaging scientific sense-making. 
\end{abstract}

\maketitle

\section{Introduction}

Modeling is a central activity in all sciences and well-known examples span many fields, including the Bohr model of the hydrogen atom, the standard model of particle physics, climate  models, and many others. Although these models apply to a diverse range of phenomena or processes, each model encodes and communicates understanding through specific diagrams, equations, and words.  Each of these scientific models abstracts and simplifies a process in the natural world by highlighting the most salient features and ignoring the extraneous details.  However, experts use models in complex ways and the process is often not made explicit in laboratory courses. We have developed a framework to describe the modeling process in physics laboratory activities. The framework attempts to abstract and simplify the complex modeling process undertaken by experts. We demonstrate that the framework captures several salient features of model-based reasoning in a way that can reveal common student difficulties and guide the development of curricula that emphasize modeling in the laboratory.  

For the last several decades, models and modeling have gained increasing attention within the science curriculum.  Although models have long been taught as content in the curriculum, only in recent decades has an explicit discussion of modeling been elevated to a central role. By 1992, David Hestenes, theoretical physicist and co-founder of the Modeling Instruction curriculum, claimed in Ref. \onlinecite{Hestenes1992} (p. 732) that, ``The great game of science is modeling the real world, and each scientific theory lays down a system of rules for playing the game.''  The 1996 National Science Education Standards (NSES) highlighted ``Evidence, Models, and Explanations'' as one of the unifying concepts and processes that all students should develop an understanding of and ability in.\cite{NSES1996}  The 2012 Framework for K-12 Science Education and the 2013 Next Generation Science Standards articulate ``Developing and Using Models'' as one of eight key science and engineering practices that should be emphasized throughout the K-12 curriculum.\cite{NGSS2013,NationalResearchCouncil2012a}  At the college level, the emphasis extends beyond the physical sciences  as well.  The 2011 AAAS Vision and Change report, which is guiding substantial innovation in undergraduate biology education, says ``Studying biological dynamics requires a greater emphasis on modeling, computation, and data analysis than ever before.''\cite{AAAS2011}  More recently the American Association of Physics Teachers Committee on Laboratories has produced a set of updated laboratory guidelines that highlight modeling as one of six central themes of the physics laboratory.\cite{AAPT2014} Although the curricular canon of core physics ideas may gradually shift, and though new subfields of science may emerge, modeling as a tool for understanding remains essential.

In this paper, we  first articulate the salient features of modeling as they are most frequently described in the physics and science education literature.  Then, we discuss limitations of prior frameworks for use in upper-division laboratory work, and describe a framework that builds on these to explicitly include modeling of measurement tools along with the physical system.  We then present results from a series of think-aloud laboratory activities where we analyze students' model-based reasoning using the framework.  Different facets of students' model-based reasoning are presented through quotes from the think-aloud activity,  and we also describe two common difficulties in modeling that were observed. Finally, we suggest future steps for investigating model-based reasoning during laboratory or experimental work.

\section{Defining models and modeling}

\begin{figure}
\includegraphics[width=0.45\textwidth, clip, trim=0mm 0mm 0mm 0mm]{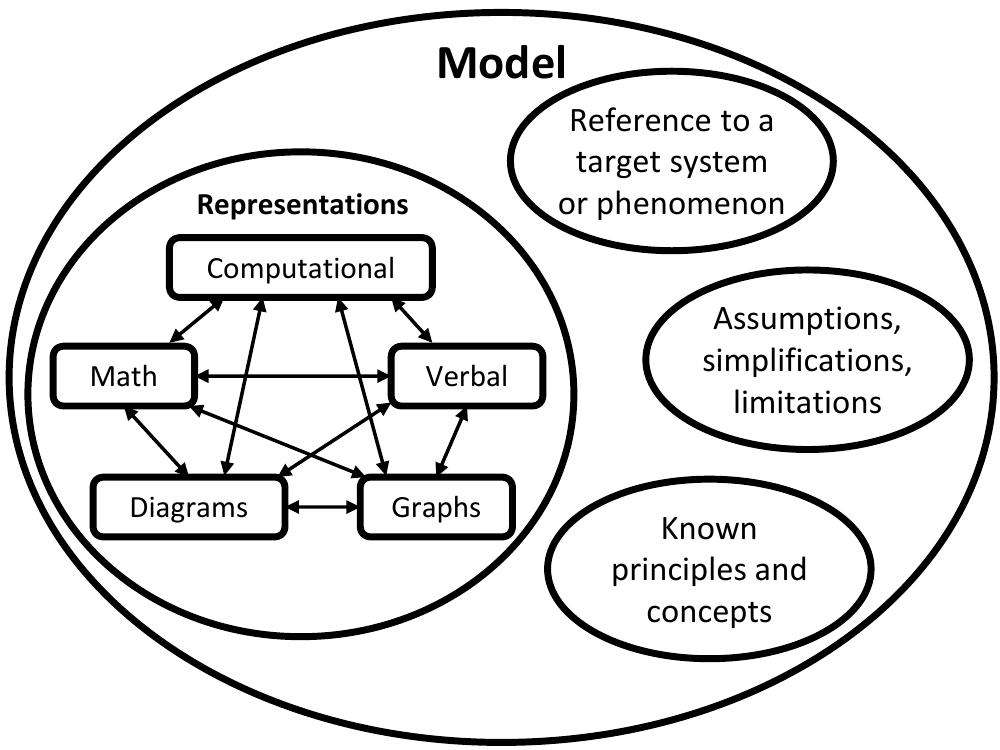}
\caption{Key components of a well-defined model include articulation of a target system, assumptions, key principles, and external representations.}
\label{fig:model_components}
\end{figure}

There is consistent agreement among several authors and national reports as to the key facets that constitute models and modeling.  Fig.\ \ref{fig:model_components} encapsulates what is included in a well-defined model.  First, a scientific model is directed at explaining or understanding some aspect of the real world.  There is a \textit{target system or phenomena} of interest.  Hestenes describes models as ``A surrogate object, a conceptual representation of a real thing.''\cite{Hestenes1987}  Or the NSES states, ``Models are tentative schemes or structures that correspond to real objects, events, or classes of events, and that have explanatory power. Models help scientists and engineers understand how things work.''\cite{NSES1996}  

Second, models are always externally articulated or \textit{represented} through words, mathematics, diagrams, and other means. A Framework for K-12 Science Education says, ``Conceptual models...are...\textit{explicit representations} that are in some ways analogous to the phenomena they represent. Conceptual models allow scientists and engineers to better visualize and understand a phenomenon under investigation or develop a possible solution to a design problem.''\cite{NationalResearchCouncil2012a}   It is equally important to note that a representation by itself does not constitute the entirety of the model.  Schwarz et al state, ``…not all representations are models. Models are specialized representations that embody aspects of mechanism, causality, or function to illustrate, explain, and predict phenomena."\cite{Schwarz2009}  Not every mathematical equation is a model, but it is typically the goal of most physics courses to demonstrate that the math is more than symbols and that equations have physical meaning and explanatory power.  In Fig.\ \ref{fig:model_components}, the arrows between the representations highlight that no single representation is sufficient, and students (and experts) must continually translate among these representations in their attempts to explain phenomena.

Third, all models are simplified and contain assumptions.  Schwarz and colleagues in the MoDeLS Project point out that a scientific model ``\textit{abstracts and simplifies} [emphasis added] a system by focusing on key features to explain and predict scientific phenomena.''\cite{Schwarz2009}  All users of a model should be aware of the simplifications and know when the model can be accurately applied to a system.   Further, because of these assumptions and simplifications,  models are tentative and undergo refinement.  Schwarz et al. summarize this aspect of modeling as ``models change as our understanding improves,'' which becomes a central dimension in their metamodeling learning progression.\cite{Schwarz2009}

The fourth component of the model as shown in Fig.\ \ref{fig:model_components} is the \textit{known principles and concepts} upon which the model is based.  Models are developed using our prior understanding of principles and concepts that may apply to the target system or phenomena.  For instance, much of Newtonian mechanics can be viewed as an application of a few principles of motion (e.g., $\vec{F} = m\vec{a}$, conservation of energy), which are used to develop mathematical representations that describe particular mechanical systems (e.g. pendulum, rolling balls, falling objects).  Halloun, another co-founder of Modeling Instruction makes this link explicit, ``A scientific model is … a conceptual system mapped, \textit{within the  context of a specific theory} [emphasis added], onto a specific pattern in the structure and/or behavior of a set of physical systems so as to reliably represent the pattern in question and serve specific functions in its regard.''\cite{Halloun2004}

One limitation of  Fig.\ \ref{fig:model_components} is that it treats the model as a static object.  Yet every model must be constructed.  Every model must be used to make predictions or explanations.  Every model should be tested empirically by evaluating the soundness of assumptions and by comparing its predictions to observations of the target phenomenon.  Finally, while a model can be well-defined, it is always tentative and subject to refinement.  As new experiments are conducted under a wider range of circumstances or with greater precision, previously satisfactory explanations no longer suffice and iterative model refinement must occur.  By \textit{modeling}, we mean this dynamic process of constructing and using models.  It is our goal here to present a framework for modeling that changes the static picture of  Fig.\ \ref{fig:model_components} into a dynamic process that is descriptive of students' model-based reasoning in the laboratory.

Although this discussion of modeling has entirely focused on models that have external representations, we acknowledge students' mental models also play an important role in learning science. The Framework for K-12 Science Education makes the distinction that ``mental models are internal, personal, idiosyncratic, incomplete, unstable, and essentially functional.'' It is the permanence of external representations that enables models to be effective tools for communication and scientific explanation, and it is these external representations that can be gradually be improved through the work of a community researchers separated in space and time.

\section{Framework for modeling in labs}
\label{sec:framework}

\subsection{Review of prior frameworks}
\label{sec:framework_prior}
One of the earlier schematics for the modeling process is a linear 4-step scheme developed by Hestenes and Halloun that includes: model description, formulation, ramification, and validation.\cite{Hestenes1987}  A later version also used by Hestenes and Halloun includes an iterative process of prediction, analysis, and validation.\cite{Halloun1996}  Both frameworks are targeted at developing models of physical systems discussed in standard lecture courses (e.g., developing models of mechanical systems in motion).  Windschitl et al. created a highly iterative framework where they reframe the entire task of inquiry-based science as centered around modeling, something they call Model-Based Inquiry.\cite{Windschitl2008} The Model-Based Inquiry framework is perhaps the most comprehensive look at a broad range of scientific practices (e.g., asking questions, building models, generating hypothesis, constructing arguments, and seeking evidence)  within a framework that emphasizes modeling  as the primary tool for scientific explanation.  Another framework from Schwarz et al. is directed at learning progressions and emphasizes the interconnectedness between metamodeling knowledge and the elements of modeling (e.g., construction, prediction, evaluation, revision), but doesn't describe how these elements form a process.  The Framework for K-12 Science Education lays out a very general schematic for how modeling fits among several other scientific practices important to scientists and engineers.\cite{NationalResearchCouncil2012a}  Finally, the prior framework that most tightly couples experiments with modeling is the learning cycle utilized throughout the introductory-level Investigative Science Learning Environment (ISLE) curriculum.  The ISLE cycle progresses through making observations, identifying patterns, creating explanations, articulating assumptions, making predictions, testing predictions, and revising models and experiments.\cite{Etkina2007a}

\subsection{The need for a upper-division lab modeling framework}
The broad challenge that inspired our new framework was the desire to describe modeling in a way that could be readily applied to  upper-division physics lab courses for analyzing student reasoning or design of curricula.  The earlier frameworks were focused on general K-12 science and introductory college physics. While there are many extremely helpful insights in each of the frameworks cited above, none of them has upper-division physics labs or experimental physics research in mind.  In the language of modeling, upper-division physics labs were outside of their domain of applicability.  Some frameworks operate at too general a level to really link to the details of the lab; some do not fully capture the iterative nature of the lab; and some do not present the connections between the elements of modeling.  The frameworks that accompany Modeling Instruction\cite{Hestenes1987, Halloun1996} are some of the most detailed and relevant to physics, yet they have a singular focus on the development of models of physical systems, and give little attention to the experimental tools that must be used to make the measurements. 

In order to demonstrate why the new framework includes measurement tools while prior frameworks were able to largely ignore them, it is worth contrasting the apparatus and measurement tools in introductory labs, where the goal is often making a quick connection to fundamental principles, with upper-division labs, which are closer to authentic physics research in their complexity and instrumentation.  Introductory labs commonly employ two techniques to quickly elucidate physical principles. First, many introductory lab experiments are engineered to make the behavior of the system agree with highly simplified models that are easily derived from principles.  The experimental details are arranged so as not to distract students' attention from the principles. For instance, the canonical ``block sliding down a plane'' is commonly implemented using a cart with wheels.  The knife edge wheels are designed to have low rolling resistance, are mounted to low friction axles, and their low mass and radius result in a small moment of inertia that ensures the rotational kinetic energy in the wheels is small compared to the translational kinetic energy of the cart. One manufacturer explicitly states that the carts are designed to ensure that ``student data more closely matches theory.'' \cite{PASCO} The second technique employed in many introductory labs is the use of sensors and automated data acquisition \cite{Sokoloff2007,Thornton1990}.  The various sensors do an excellent job of quickly obtaining quantitative data and providing real-time graphic visualizations of the data. With such tools, students can quickly compare predictions to measurements, make refinements to their models, and run a test again.  However, the principles of operation and performance limitations of the measurement tools are rarely discussed.  The measurement tools are engineered so that any deviation from ideal measurement operation can be ignored when used for common introductory experiments.  Further, the model of the measurement tool is often integrated into the apparatus and software, so every sensor directly outputs the quantity of interest.  For example, force sensors output newtons while position sensors output meters even when those quantities are not directly measured (e.g., a motion detector converts a time delay between emitted and received pulses into a distance). Students are encouraged to focus on the relationship between force and motion rather than on how these measurements work or on statistical uncertainty and systematic errors of the sensors.

While prioritizing basic physics principles is natural for an introductory course, upper-division physics majors need a more holistic view of the experimental process.  This holistic view should include a deep understanding of the principles of physics underlying the entire apparatus, including the measurement tools.  In a recent survey about laboratory learning goals, faculty frequently mentioned the importance of understanding the ``black boxes'' in the laboratory through an understanding of the principles and limitations of operation.\cite{Zwickl2013} Modeling the full apparatus affords an important professional development opportunity for students pursuing careers in physical science or engineering.  Physicists, as a community, are frequently pushing the limits of commercially available measurement tools and are often on the leading edge of designing new measurement techniques (e.g., Magnetic Resonance Imaging), which is only possible through a deep intellectual engagement with the entire experiment.  Thus, what is needed is a extension of existing modeling frameworks \cite{Hestenes1992,Schwarz2009} that is more applicable to the complex apparatus of upper-division labs.

\subsection{A framework for modeling in upper-division labs}
As described earlier in Sec.\ \ref{sec:framework_prior}, most frameworks for describing modeling at the introductory level emphasize the development of new principles and concepts, while the measurement process is largely taken for granted. Hestenes does say, ``...there is a critical theoretical component to the design of every piece of apparatus and everything [experimentalists] do in the laboratory'' and later quotes Martin Deutsch, ``How it is possible that important and reliable conclusions are drawn from this experimentation? The answer lies in the fact that the experimenter starts out with a well structured image of the actual connections between the events [of turning a knob and the recorded measurement].''\cite{Hestenes1992} Despite this acknowledgment, the modeling frameworks don't explicitly include measurement tools.  Beyond the omission of measurement tools, most upper-division physics labs do not seek to inductively develop new fundamental principles (e.g., Maxwell's equations), but more commonly to apply known principles to explain observable phenomena or test predictions.  Although the underlying physics principles are typically assumed to be prior knowledge for the students, there is still substantial opportunity for constructing models, identifying key principles and parameters, making relevant assumptions, and making predictions that are testable.  Because many upper-division labs involve advanced concepts and use equipment that is often unfamiliar, students must do significant intellectual work in order to understand the experiment.

\begin{figure}
\includegraphics[width=0.8\textwidth, clip, trim=0mm 0mm 0mm 0mm]{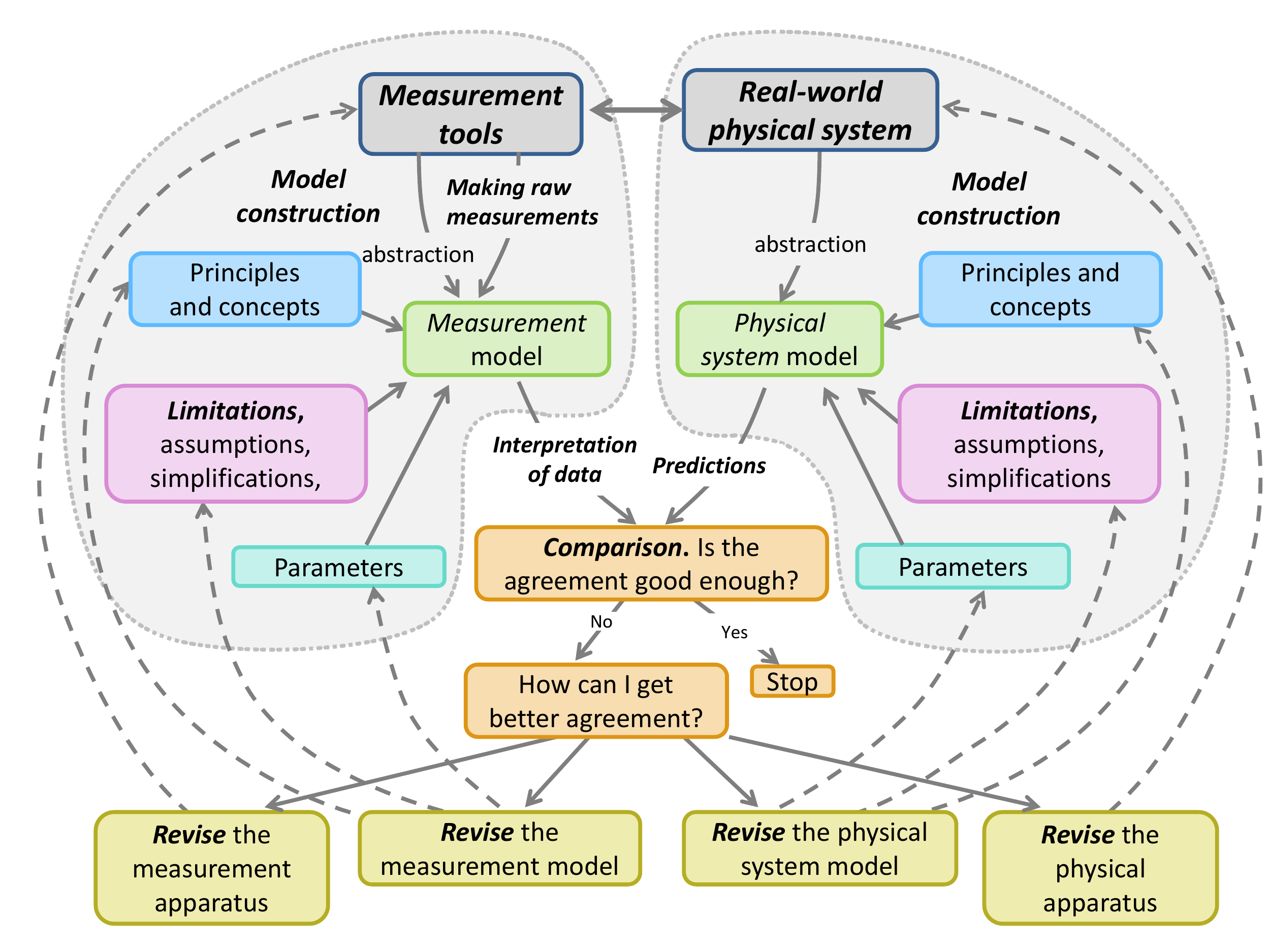}
\caption{The modeling framework describes a process that includes constructing models of the measurement tools and physical system, making predictions, making measurements, interpreting measurements using the measurement model, comparison between predictions and measurements, and several pathways for revision of the models and apparatus. The labels in italics correspond to key facets of modeling that are also coded in the think-aloud laboratory activities (see Figs.\ \ref{fig:Meas_vs_phys} and \ref{fig:Modeling_codes}).  The \textit{model construction} phases shown in the top right and top left aligns with the key components of models shown in Fig.\ \ref{fig:model_components}  Color is used in the diagram to highlight the distinct phases of the modeling process and to highlight the symmetry between the modeling process for the measurement tools and physical system.}
\label{fig:Modeling_framework}
\end{figure}

The holistic framework for modeling in the laboratory in Fig.\ \ref{fig:Modeling_framework} divides the full experimental apparatus into the \textit{physical system} (right side of Fig.\ \ref{fig:Modeling_framework})  and the \textit{measurement tools} (left side of Fig.\ \ref{fig:Modeling_framework}). (Italics are used throughout the paper to highlight the connections to key stages in the framework in Fig.\ \ref{fig:Modeling_framework} and in the modeling codes used in Figs.\ \ref{fig:Meas_vs_phys} and \ref{fig:Modeling_codes} and Table\ \ref{tab:codes}.) This conceptual division between measurement tools and physical system is useful because it recognizes that models used to understand the physical system (e.g., a block sliding down a ramp) are quite different from those used to understand the measurement tools (e.g., a motion detector). Sometimes the division is obvious in an experiment, while other times, the measurement tools and physical system are more integrated.  In such cases, while there might be multiple ways to conceptually divide the system into a physical system and measurement tools, any reasonable division is still a helpful for modeling the full apparatus.  The left-right symmetry of the framework emphasizes that the measurement tools, in addition to the physical system, can be understood through mechanisms rooted in principles of physics. 

To begin working through the modeling diagram, we will start at the model \textit{construction} phase,\footnote{Italics denote aspects of the modeling framework that also serve as codes in the later analysis of student work.} depicted in the top left of Fig.\ \ref{fig:Modeling_framework} (for measurement tools) and top right (for the physical system). Following the description of key model components in Fig.\ \ref{fig:model_components}, a model is an abstraction of some real device, and the construction requires an input of principles and concepts, \textit{limitations} and assumptions, and key parameters. Although the construction stage is important for both the measurement tools model and the physical system model, there are typically differences in the details of the construction process. The construction of the physical system model typically builds heavily on core ideas from undergraduate courses, while the principles, limitations, and parameters for the measurement tools model are largely supplied through manufacturers' documentation. Reading a data sheet is reframed as an exercise in model-based reasoning. Regarding use, the physical system model is used to make \textit{predictions} (a stage represented in other modeling frameworks), but the measurement tools model is used to \textit{interpret raw data} or translate raw observations into physically meaningful quantities (e.g., a motion detector converts a time delay between sent and received pulses into a position). The predictions and interpreted measurements converge at a \textit{comparison}, and the results of that comparison are either satisfactory, or prompt a revision in the models or apparatus.  The framework identifies four major pathways for \textit{revision}: refine the measurement tools apparatus, refine the measurement tools model (e.g., calibrate the measurement tools), refine the physical system apparatus, or refine the physical system model.  Depending on likely sources of discrepancies (e.g., limitations were identified in the model construction phase, or a parameter was unknown in the apparatus) and upon which revisions are most easy to implement, the various pathways can be prioritized and explored.  The process is represented as a cycle that can be repeated until the experimental goal is met.

\subsection{Application of the framework to lab design}
In order to briefly demonstrate the general utility and detail included in the modeling framework in Fig. \ref{fig:Modeling_framework}, we first apply it to analyze the structure of an introductory lab activity and then apply it as a curriculum design tool that can guide the creation of activities that emphasize different aspects of modeling. For a more in-depth discussion of the framework applied to upper-division laboratory design, see Ref. \onlinecite{Zwickl2014b}, where the framework was used to develop a lab activity on the polarization of light.

To demonstrate the application of the framework in the analysis of modeling in an activity, consider a standard lab activity where students construct a simple pendulum of length $L$ and measure it's period $T$ for small angle oscillations to determine the gravitational acceleration $g$ using a relationship previously derived in class ($T = 2\pi\sqrt{L/g}$). Applying the framework to this hypothetical activity, we identify the physical system as the pendulum, while the measurement tool is a stop watch.  The model \textit{construction} stage for the physical system was mostly accomplished prior to the lab, perhaps by the instructor or textbook.  The relevant principles and simplifications were already applied in the derivation of the period. The only missing element in the model are two parameters: $L$, the pendulum length, which may be chosen by the student, and $g$, which is treated as an unknown model parameter.  The measurement tools produce \textit{raw data}, which are the total time elapsed on the stop watch $\Delta t$ and the number of oscillations $N$.  The \textit{measurement tools model} is very simple and has a mathematical representation as $T=\Delta t/N$. The use of the measurement tools model gives the \textit{interpreted data}, which is the period per oscillation. Viewed as a modeling exercise, this ``measurement of $g$'' lab is an exercise where Newton's laws of motion are known, the gravitational force has been mathematically modeled, but we imagine that within this theory of the gravitational force, there is an unknown parameter $g$, which must be found.  There is a small degree of model \textit{refinement} in the activity, in the sense that the previously ``unknown'' parameter $g$ is optimally chosen to produce the best agreement between the predicted period and the measured period.  Although this activity has some elements of modeling, it only explores a limited subset of the modeling process.

As a curriculum design tool, the framework offers a way to generate various alternative pendulum labs that emphasize additional aspects of the modeling process that go beyond measuring a particular parameter or physical constant. Based on the different stages of the framework, the lab could: (a)  Emphasize model \textit{construction}. Students could apply principles to develop the model of the pendulum if it has not been covered in class already. (b) Emphasize \textit{limitations}, simplifications, and assumptions of the pendulum and measurement tools. Students could articulate all the assumptions in the simple pendulum (e.g., small angle, point mass, etc.), describe the limits of validity for each assumption, and justify whether or not their pendulum experiment satisfies those conditions. Students could then test the gradual breakdown of the model as the real pendulum is designed and operated in a regime that goes beyond the model's limits of validity.  (c) Emphasize testing a broader set of \textit{predictions} of the simple pendulum theory, such as, $T\propto\sqrt{L}$, $T\propto1/\sqrt{g}$, or that $T$ does not depend on mass $m$ or amplitude of oscillation.  Students could then design experiments to test each of these predictions. (d) Emphasize \textit{limitations} and \textit{refinement} of measurement tools. For instance, video analysis of an oscillating pendulum (e.g., using Vernier Logger Pro or Tracker Video Analysis and Modeling Tool\footnote{Tracker Video Analysis and Modeling Tool, https://www.cabrillo.edu/$\sim$dbrown/tracker/}) offers several opportunities to discuss limitations and refinement of measurement models as students' discover the need to keep their camera at a fixed position and orientation, choose an appropriate camera angle, and understand the geometric distortion of the images due to the lens.  While these alternatives don't constitute an exhaustive list of possibilities, they do exemplify how the framework can inspire a broad range of modeling activities by repurposing standard lab equipment.

\section{Think-aloud lab activities}
\label{sec:interviews}

In addition to using the framework for curriculum design, we also use the framework to measure students' engagement in different aspects of the modeling process in the laboratory.  Because one new and significant component of our framework was the emphasis on modeling the measurement tools, we wanted to find examples of how students engage in a modeling process with their measurement tools during laboratory activities (Sec.\ \ref{sec:student_excerpts}). Further, we were curious if students found particular aspects of the modeling process to be especially challenging (Secs. \ref{sec:assumptions} and \ref{sec:insufficient_concepts}).  

\begin{figure}
\includegraphics[width=0.45\textwidth, clip, trim=0mm 0mm 0mm 0mm]{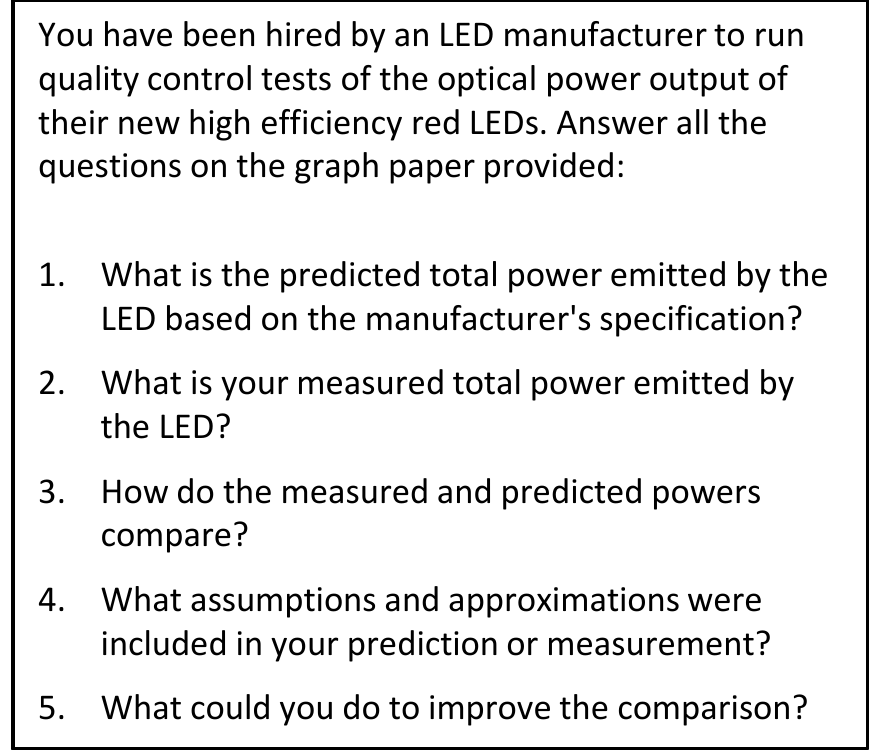}
\caption{Written prompts used for the think-aloud experimental activity.}
\label{fig:activity_sheet}
\end{figure}

\begin{figure}
\includegraphics[width=0.45\textwidth, clip, trim=0mm 0mm 0mm 0mm]{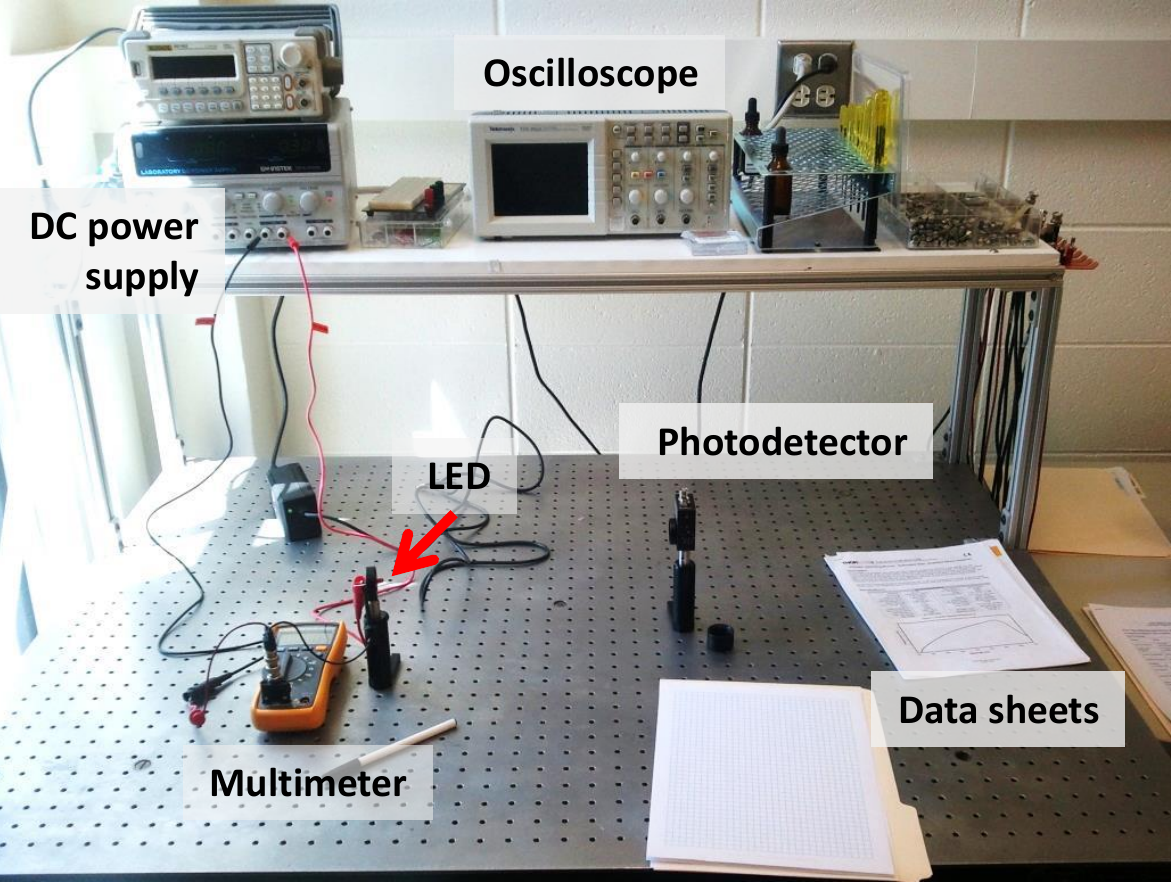}
\caption{Laboratory bench used for the think-aloud experimental activity.}
\label{fig:Lab_bench}
\end{figure}

In order to capture students' experimental work involving modeling, we designed a short laboratory activity that would provide students with ample opportunities to design, use, test, and refine models of the physical system and measurement tools. The think-aloud activity was based on an instructional lab that students had previously completed in a junior-level electronics course. Students were given an LED connected to a DC power supply and asked to measure the LED's optical power output and compare that to a value they predict using information from a product data sheet. The short prompt for the activity is shown in Fig.\ \ref{fig:activity_sheet}, and Fig.\ \ref{fig:Lab_bench} shows the lab bench and equipment that students were able to use during the activity. The experiment was designed to provide students opportunities to construct and refine models of the measurement tools and the LED with varying assumptions and levels of sophistication.

The \textit{physical system} consisted of a red LED connected to a DC power supply in series with a resistor to limit the maximum current through the LED.  A multimeter was used to measure current through the LED.  The \textit{measurement tools} used to determine the optical power output were a Thorlabs PDA36A Adjustable Gain Photodetector and an oscilloscope.  Students were also provided with data sheets for the LED, photodetector, multimeter, and oscilloscope along with assorted optical mounts.

In order to elucidate the design of the LED activity, we describe one possible approach that would lead to an accurate prediction and measurement of the LED's optical power output.  The first step,  predicting the optical power output, requires knowledge of the typical angular intensity (given on the LED data sheet as $I_\text{ang} = 670$\ \textmu W/steradian) and an estimate of the solid angle of the cone of light emitted by the LED. The width of the cone was specified as $\theta  = 22\degree$ (0.38 rad). Assuming the cone width is small ($\theta <<\pi$),  the area $A$ illuminated by the cone at a distance, $R$, from the LED is $A\approx  \pi(R \theta/2)^2$.  The solid angle subtended by the cone is $\Omega = A/R^2 = \pi (\theta /2)^2 \approx 0.11$ steradians.  The predicted optical output power is $P = I_\text{ang}\Omega \approx 76$ \textmu W.  The second part, measuring the optical power output, requires setting the LED current to the manufacturer's test condition (20 mA) and positioning the photodetector extremely close to the LED so as to capture the entire cone of light in one measurement.  The measured voltage from the photodetector can be converted into incident optical power using the known gain setting on the detector and the wavelength of the emitted light (LED has a peak wavelength at 635 nm and and a spectral width of 45 nm). Finally, there may be an offset voltage in the photodetector output due to ambient light in the room or the internal operation of the detector, which should be subtracted from the measurements of the LED power output.  This offset voltage can be determined by measuring the photodetector output voltage with the LED turned off.   

The student participants were drawn from an upper-division optics and modern physics lab course for physics majors at a large public research university. The course has a typical enrollment of 20-25 students.  All students enrolled in the course were invited to participate in the think-aloud interview, and eight students responded. All eight participants were male, which reflected the fact that 20 of 21 students enrolled in the course during the semester of data collection were male.  As an upper-division lab course where nearly all students complete their assignments, course grades are typically split between A's and B's.  Of the eight students who were interviewed, four received A's and four received B's.  Of the three students analyzed in detail in Sec. \ref{sec:Results} and Figs.\ \ref{fig:Meas_vs_phys} and \ref{fig:Modeling_codes}, two received an A and one received a B. The participants received a \$20 incentive for their time and effort, but participants did not earn any credit or extra credit in the lab course.  The particular sampling of students was not chosen in order to make generalizable claims about specific patterns in student reasoning, but rather to provide evidence that the new framework for modeling offers significant utility for capturing and differentiating student reasoning during a complex lab activity. 

The interviews were conducted during the final two weeks of the semester, by which time the lab space and equipment shown in Fig.\ \ref{fig:Lab_bench} (other than the LED) were part of their weekly lab experience. The think-aloud lab portion of the interview typically took 30-45 minutes. After the activity, there was a 20-30 minute follow-up discussion where the students' models were reviewed in order to clarify missing details from the think-aloud portion. At the very end, the participants were allowed to request explanations for any of the physical or mathematical details of the experiment.

Audio and video of the think-aloud interviews were collected along with written observations and copies of students' work.  The full interviews were transcribed, but only the think-aloud laboratory portion was coded for instances of model-related activity identified in the framework. Table \ref{tab:codes} provides a list of high-level modeling codes and a brief description of the coding criteria. Several of the codes (\textit{model construction}, \textit{predictions}, \textit{comparison}, \textit{limitations}, \textit{revision}) are common to most frameworks that discuss modeling. Several of the high-level codes had more refined sub-codes (e.g., revision had sub-codes to indicate whether apparatus or models were revised), but these are not described in Table\ \ref{tab:codes} or shown in the data.  Several other modeling codes in Table\ \ref{tab:codes} are unique to the framework: \textit{physical system}, \textit{measurement tools}, \textit{interpretation of data}, and \textit{making measurements}.  Finally, some significant moments of student reasoning were placed into categories that were outside the framework.  \textit{Troubleshooting} was coded whenever a student recognized a problem with the apparatus and attempted to solve it. Sometimes the resolution was simple and unexpected (pressing the ``autoset'' button on the oscilloscope), while other times students engaged in what appeared to be a rapid modeling cycle involving a series of qualitative predictions and qualitative measurements (e.g., if I put my hand in front of the detector, the photodetector output voltage should decrease) in order to identify the source of the problem. In these model-based troubleshooting episodes, the timespan for a series of predictions, measurements, comparison, and refinement could occur in the span of a minute or less.  Although not all troubleshooting utilized model-based reasoning, it was remarkable to see how the process of identifying and solving the problem became a genuine episode of scientific inquiry activity where the solution was unknown to the student and yet of the utmost importance to find.

The three interviews that are included in this analysis (see Figs.\ \ref{fig:Meas_vs_phys}, \ref{fig:Modeling_codes}) were independently coded by two of us and then compared for consistency. As the coding criteria were refined, the two coders were able to come to an agreement level exceeding 95\% for the high-level modeling codes presented.

\begin{table}
\begin{center} 
    \begin{tabular}{| p{3cm} | p{13cm} |}
    \hline
    \textbf{Code} & \textbf{Criteria}  \\ \hline
    Physical System & Any activity, discussion, or modeling related to the LED and the power supply that provides current to the LED.  \\ \hline
    Measurement Tools & Any activity, discussion, or modeling activity related to the photodetector and the oscilloscope.  \\ \hline
    Construction  &   Student identifies principles, assumptions, parameters that go into the models of physical system or measurement tools. Includes constructing verbal representations (typically a conceptual discussion) and mathematical representations of the model.\\ \hline
    Prediction & Student uses the physical system model to predict the power output of the LED.    \\ \hline
    Interpretation of Data  &  Student uses the measurement tools  model to convert raw (or direct) measurements (i.e., voltage)  into a more useful quantity (i.e., optical power) so they can be compared with something else, typically a prediction.\\ \hline
    Making Measurements  &  Student gets raw results from measurement tools (primarily info from the oscilloscope, multimeter, or an observation of the phenomena made with the naked eye).  \\ \hline
    Comparison  &  Student makes a comparison between a prediction and measurement or with some prior knowledge \\ \hline
    Limitation  & Student identifies a non-ideal feature of the experimental setup or model.  Includes making assumptions and justifying assumptions.  \\ \hline
    Revision  & Student modifies some aspect of the the apparatus or models related to the physical system or measurement tool.   \\ \hline
    Troubleshooting  & Student recognizes there is a problem (some aspect of the apparatus of measurement does not work right or does not work at all) with an unidentified source and the student tries to fix it.  Typically involves a series of qualitative predictions and simple qualitative experimental tests to identify the source of the problem. \\
    \hline
    \end{tabular}
\end{center}
\caption{High-level modeling codes and brief descriptions of the coding criteria.}
\label{tab:codes}
\end{table}

\section{Results}
\label{sec:Results}
There are three main results in the sections that follow.  The first, and most important result, is that students do engage in meaningful modeling of the measurement tools. Students' modeling is significant in both duration and in quality as Figs.\ \ref{fig:Meas_vs_phys} and \ref{fig:Modeling_codes} and several interview excerpts will demonstrate.  The second and third results are about common challenges that students' had during the modeling process.  The second result reviews challenges in identifying assumptions during model construction and in justifying those assumptions and connecting to limitations.  The third result describes difficulties students had during model construction because of they had insufficient understanding of key concepts in the experiment.  This result emphasizes a link between students' conceptual understanding and their engagement scientific practices such as modeling and experimental design.

\subsection{Students'  modeling of measurement tools}
\label{sec:student_excerpts}

\begin{figure}
\includegraphics[width=0.55\textwidth, clip, trim=0mm 0mm 0mm 0mm]{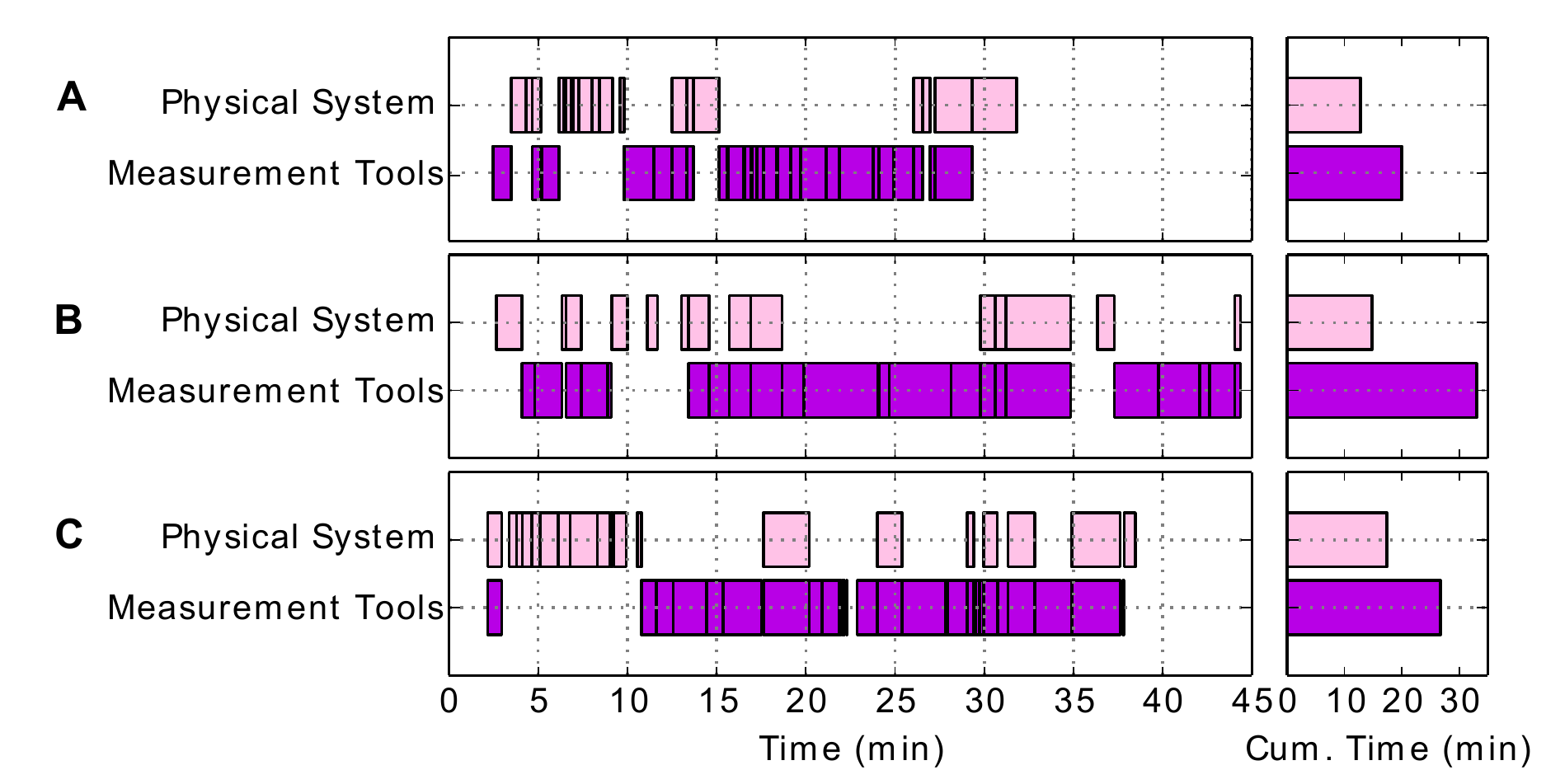}
\caption{Comparisons of students activity related to the measurement tools vs physical system for three students.  Each of the three students spent more time using and modeling the measurement tools.}
\label{fig:Meas_vs_phys}
\end{figure}

\begin{figure}
\includegraphics[width=0.95\textwidth, clip, trim=0mm 0mm 0mm 0mm]{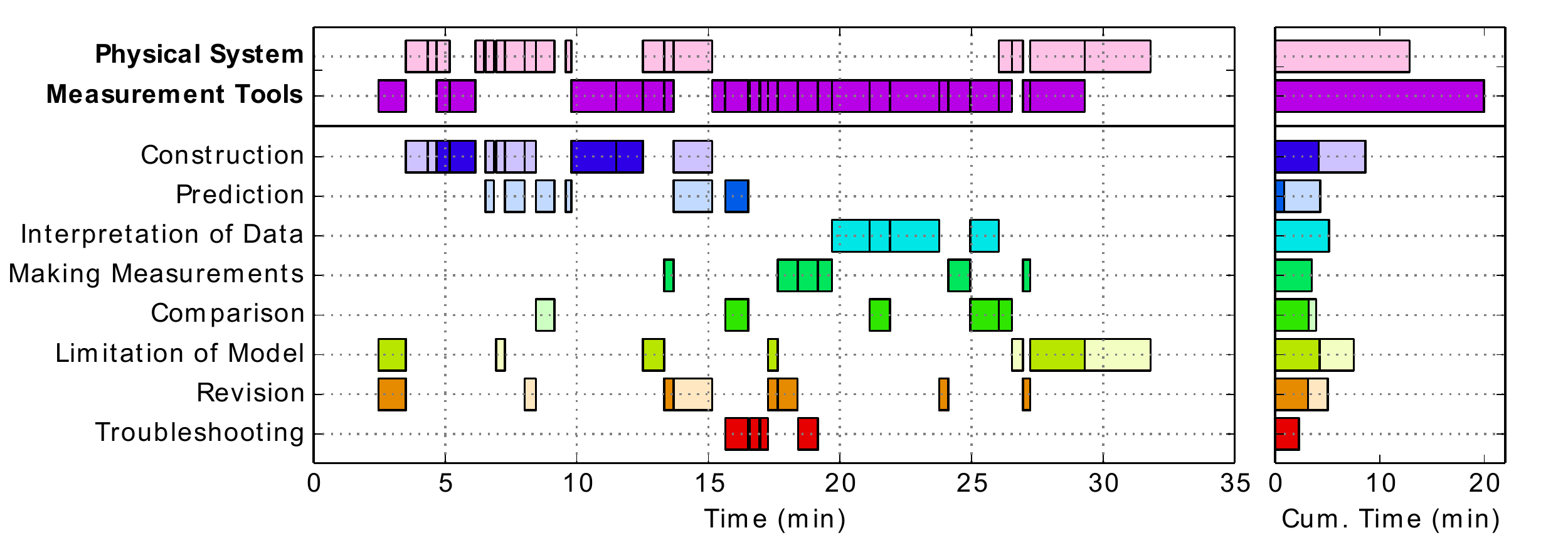}
\caption{Detailed view of Student A's modeling codes throughout the laboratory activity.  Darker shades are modeling activities specific to the measurement tools.  Lighter shades are modeling activities related to the physical system.}
\label{fig:Modeling_codes}
\end{figure}

Fig.\ \ref{fig:Meas_vs_phys} shows the occurrences of physical system (LED) and measurement tools (photodetector and oscilloscope) codes throughout the interview.  As Fig.\ \ref{fig:Meas_vs_phys} shows, each of the students spent a significant fraction of their time using and modeling the measurement tools in this particular upper-division lab activity. Based on Fig.\ \ref{fig:Meas_vs_phys} alone, it would be impossible to understand the character or quality of the measurement tools activity because it may just be that the students were mindlessly turning knobs and writing down measurements.  We demonstrate the quality in two ways.  First, Fig.\ \ref{fig:Modeling_codes} shows the occurrences of specific modeling codes for Student A over the entire think-aloud activity.  The \textit{making measurements} code does specifically code the activity of turning knobs and making observations, but Fig.\ \ref{fig:Modeling_codes} shows this only occupies about 4 minutes of the activity, while a broad range of other modeling activities occur for comparable lengths of time throughout the interview  (e.g., construction, interpretation of data, comparison, revision).  Although the pattern of codes does vary for each student, Fig.\ \ref{fig:Modeling_codes} is representative in the frequency of different modeling activities. Second, we use several interview excerpts to show of the quality of students' model-based reasoning.  Examples are provided that span four different aspects of modeling in relation to the measurement tools:  construction, identification of assumptions and limitations, interpretation of data, and revision. The primary claim, which follows from Figs.\ \ref{fig:Meas_vs_phys} and \ref{fig:Modeling_codes} and the transcript excerpts below, is that frameworks for modeling in the lab must explicitly include the use of measurement tools. If a modeling framework does not explicitly include the measurement tools, it may overlook frequently occurring episodes of model-based reasoning, which are especially likely to occur in upper-division physics labs that involve sophisticated measurement equipment.

\textbf{Excerpt: Model construction---Principles and concepts.} One element of model construction is identifying the key principles and concepts in the model. In this excerpt, Student A is trying to understand the physical mechanism by which the incident light on the detector is converted to the output voltage $V$ being measured by the oscilloscope.  The student realizes that the intensity of the broadly dispersed light from the LED must be integrated over some area to get a total incident power $P_{in}$.  The student correctly identifies that the power from the absorbed light produces a current in the photodiode (according to a material-dependent proportionality constant $R$, known as the responsivity).  The current is converted to a voltage using a circuit known as a transimpedance amplifier with gain $G$ in volts per amp.  Mathematically, this model is represented as $V = GRP_{in}$.  

\textit{``I've got to figure out how to turn voltage measurements into power.  I guess I don't have any idea how to do that honestly.} (Student is reading data sheet) \textit{Tell me power to voltage...voltage to power... spectral responsivity.  I kind of remember this stuff.   Something like when light [is] incident on the photo detector and that intensity is integrated into a power and that power causes a current and the current transimpedance gain...  So I have voltage and I'm looking for a power . This is figuring out power from a measured voltage.''} The student goes on to successfully construct a quantitative model connecting a measured voltage to an incident power. However, by first identifying the principles of operation, the mathematical equation that is used is not just a computational tool, but has been clearly linked to the physical setup.

\textbf{Excerpt: Model construction---Limitations.} A second element of model construction is identifying the limitations in the models and apparatus.  Student B recognizes and addresses two limitations. First, the LED emits light over a range of angles, but the photodetector's small size may not be large enough to capture the entire cone of light.  The effect of the limitation can be minimized by bringing the detector as close as possible to the LED.  Second, the student observes that ambient light is also incident on the detector, the effect of this limitation can be reduced by blocking more ambient light. 

\textit{``I think I'm just gonna put the photodiode directly in front of the LED and hope that it's in between the 23 degrees of the viewing angle so I get approximately 100 percent of the relative intensity...I just put the photodiode close so, that makes sure that I get as much of the intensity as I can. I'm gonna put this [light blocking tube] on to eliminate a lot of the sunlight.''} 

\textbf{Excerpt: Interpretation of data.} Student B constructs a model of the measurement tools, complete with actual device parameters for the peak wavelength of the LED, the responsivity of the photodetector, and the transimpedance amplifier gain.   The student then uses this mathematical model with parameters to obtain the desired measurement (optical power output of the LED) from the raw voltage measurements.

\textit{``I know the voltage is proportional to the intensity. I know the photodiode has got a small diode in there and it's only picking up a small area of intensity from the LED so, that's the area, if I knew that area and multiplied it by I, I'd get watts. Power equals $I$ times $A$. I'm still struggling to figure out how I'll get voltage into intensity. Once I have intensity, I can get the power. I know volts are watts per amp...I know I need to use this responsivity which is an amp per watt in the peak wavelength for the LEDs is 635 [nm], so 635 on this gives me a responsivity of 0.4 amps per watts. So I need to get this into power. So, one over responsivity gives me watts per amp so, I need to calculate the current from the photodiode.  I'm gonna use the gain which is at 0 dB setting so the gain is volts per amp which I believe is $1.51\times10^3$ volts per amp. That times the responsivity of 0.4 gives us volts per watt which is 604 volts per watt, so one over that gives me the wattages per amp er... per volt. ...So, now multiplying our $\Delta V$ which is 22.2 mV times this watts per volt gives us the intensity as seen in the photodiode. I get a power of $3.84\times 10^{-5}$ W. That was the power, the measured power from the photodiode.''} 

\textbf{Excerpt: Revision of apparatus and model.} Student B revises the apparatus to minimize ambient light and revises the model of the photodetector to include an offset in the photodetector output voltage that is present even when the LED is off (i.e., $V = GRP_{in}+V_{off}$). It is interesting to see the use of two simultaneous model revision strategies, which represent two of the four pathways shown in the revision stage of Fig.\ \ref{fig:Modeling_framework}---revising the measurement apparatus to be more ideal (lower the offset due to background light) and generalize the measurement tools model to account for remaining offset voltage $V_{off}$.  

\textit{``It looks like the photodiode is definitely detecting [light]. Okay the photo diode is detecting a lot of sunlight. It's detecting the diode but the sunlight is definitely affecting the reading I'm just trying to think of the best way to get it. I think I'm just gonna put the diode right up against this thing to minimize all the stray light.It looks like with the LED off there's an offset voltage of just under 20 mV. ...With the LED on, the voltage goes up to 39.2 mV. ...The difference is 23.2 mV from the diode being on to off. Now, we're gonna turn the voltage output into total power emitted by the LED.''}  

Both the quantitative and qualitative data support the idea that modeling frameworks applicable to upper-division physics labs must include modeling of the measurement tools.  The ratio of students' effort devoted to modeling the physical system versus measurement tools will almost certainly vary depending on the particular laboratory activity. The particular balance shown in Figs.\ \ref{fig:Meas_vs_phys} and \ref{fig:Modeling_codes} is due in part to the intentional use of a common measurement tool (the photodetector), which requires modeling as a natural part of gathering and analyzing data.  However, most upper-division laboratories include a wide variety of measurement tools and techniques for which this modeling framework will be well-suited.

\subsection{Students' challenges in model construction when articulating assumptions}
\label{sec:assumptions}
Although articulating assumptions is a common element of modeling in nearly all descriptions of model construction, we found several instances during the think-alound interviews where students would utilize a model, but not recognize the assumptions that supported that model. This difficulty is significant because it may hinder other aspects of the modeling process.  First, an unidentified assumption is not going to be justified, and so will be included without any critical evaluation as to its appropriateness. Second, the assumptions will not be connected to limitations of the model. Finally, those unidentified limitations are unlikely to inspire any iterative refinements to the experiment. 

A common occurrence in the think-aloud activity was that students predicted the optical power using $P=IV$ where $V$ is the voltage drop across the LED and $I$ is the current through the LED. Students identified the relevant parameters in the LED data sheet and computed a numerical result for the power $P=IV$.  For example, Student B said, \textit{``I need to predict that total power of the LED.  And so the power is the current times the voltage.  And the forward voltage drop is like 2 volts . . .well, I'll try and put 20 milliamps in it because that's what it told me to do... which means my power should be 2 volts times .02 is 0.04 watts. So predicted power is 0.04 Watts, so good stuff.''} The student used a particular principle from electronic circuits ($P=IV$), and identified relevant parameters in the device, yet did not recognize that the model assumes 100\% of the electrical energy used would be converted into light. Because the assumption was not identified, no attempt was made to justify its appropriateness, and a modification of the assumption was never an option for model revision.  Toward the end of the interview, after the think-aloud portion was complete, the student was directly asked \textit{``So you calculated [power] based on the voltage drop across the diode, the current running through the diode. So, what assumption were you making about the optical power output?''} The student replied, \textit{``I guess that the optical power output would be the same as the power used in the circuit like whatever power was in the circuit was all emitted light, which isn't necessarily the case I guess.''}  In this case, with a direct question, the student did reflect upon the LED model and recognize the assumption and that it didn't have any justification.  A general pattern was that when a mathematical representation for the model could be readily identified (e.g., $P=IV$), then an explicit discussion of the assumptions was bypassed.  

A second example was in the construction of the measurement model involving the conversion of optical power $P$ into photodetector output voltage $V$. A mathematical relationship was provided in the photodetector data sheet: $V = GRP_{in}$, where the responsivity $R$ depends on the material (in this case silicon) and the wavelength of the light. However, the LED emitted a spectrum of wavelengths with a spectral width of about 45 nm. When listing the various assumptions that were made as part of the prediction or measurement, none of the students listed the assumption that the light was monochromatic, although many other assumptions were listed. When specifically asked if there were any assumptions about the spectral properties of light, all students immediately responded that they assumed the wavelength of the LED was a single wavelength at the peak of the spectrum shown on the data sheet.  One of the students went on to justify this assumption: \textit{``I would expect that half of the wavelengths [in the spectrum] would give less responsivity  and half would give more because we are kind of in this spot where [the responsivity] is almost linear but it does vary... So half would be less [and] half would be more responsivity which would lead to hopefully the same calculated [value as when assuming a monochromatic source].''}  In this case, the student was able to provide a justification, but only when the assumption was brought to the student's attention. 


\subsection{Students' challenges in model construction from insufficient conceptual understanding}
\label{sec:insufficient_concepts}
In addition to recognizing assumptions, there is a certain amount of prior knowledge that is needed for the construction of a model. In the think-aloud activity, the most accurate model for predicting the optical power output of the LED required the use of an angular distribution of power in microwatts per steradian. Because LEDs are designed to have a particular emission pattern for the light output depending on their application, the data sheet provided a polar coordinate plot of the relative intensity as a function of angle. Also, the data sheet specified the numerical value of the maximum power output per unit solid angle (in microwatts per steradian) and the approximate angular width of the emission pattern. We anticipated students would be able to estimate the emitted power by roughly determining the solid angle of the emitted cone of light and multiplying that solid angle by the peak angular intensity.  

The following brief explanatory comment about solid angle was also provided on the data sheet: ``A steradian is a measure of solid angle, which is the area of the angular region on a unit sphere.  A full sphere subtends $4\pi$ steradians.''  Despite the explanatory comment and the fact that a similar activity in a prerequisite lab course used solid angle, most students were unsure about the meaning of steradians as a unit.  Within the interviews, the related concept of intensity as power per unit area was used by the students.  However, the conceptual modification of angular intensity as power per solid angle produced significant confusion in the think-aloud activity. 

After Student A completed the entire think-aloud activity, the interviewer mentioned the student had omitted any reference to the use of angular intensity in microwatts/steradian on the data sheet.  When asked \textit{``What do you make of this angular intensity spec?''} Student A replied that, \textit{``I guess I kind of glossed over that.''} Student A's response could have indicated a simple oversight, but only after a 10 minute back-and-forth discussion was the student able to get a rough estimate of the solid angle of the LED emission pattern and use it to produce a new prediction for the optical output power.  This indicates that the student's oversight was likely connected to a lack of familiarity with the concept of solid angle and the unit of steradian. 

Other students also had difficulty with the concept of solid angle.  In the process of making predictions, Student B said \textit{``I am calculating the power $IV$ for the diode according to the data sheet. I'm not sure what to do with the angular intensity. I don't know the units of microwatts per steradian.''}  Student C, when first reading through the provided data sheet, expressed \textit{``What is that kind of graph? Relative intensity...I have no idea what that is.''}

Each of these students ended up utilizing a $P=IV$ model for optical power output because it was the most obvious model that avoided the concepts of solid angle and steradians.  Further consultation with undergraduate instructors confirmed that the one-week lab done in the earlier electronics lab course was probably the only time in their core curriculum where the concept of solid angle was directly addressed.

When a lab activity utilizes concepts that are largely outside of students' prior knowledge, it has a significant impact on how they engage in the laboratory.  In this case, it caused most students to largely ignore key information in the data sheet, and construct significantly less accurate models of the optical output power (the $P=IV$ model produced a prediction about 500 times larger than using angular intensity in the model). Beyond model construction and predictions, the angular emission pattern played an additional role in designing the experiment, as it provides one justification for whether or not the photodetector could capture the full emission pattern of the LED.  

\section{Conclusions}

Reviewing the prior literature on modeling suggests most earlier modeling frameworks emphasize modeling physical systems that closely relate to core physics ideas (e.g., mechanics or electricity and magnetism).  In these frameworks, much of the experimental apparatus is largely overlooked in the modeling process in the pursuit of key principles and concepts.  However, in upper-division physics laboratory courses, there is often a non-trivial relationship between the phenomena being studied and the raw measurements, which require students to have a sophisticated understanding of the design and operation of the full experimental apparatus. 

This paper presents a new framework for modeling that treats the full experimental apparatus in two parts: the measurement tools and the physical system, and both parts are subject to a modeling process as a way to understand the design and operation of the experiment. The framework serves as both a descriptive tool for characterizing students' model-based reasoning and has applications as a curriculum design tool. Analysis of multiple think-aloud experimental activities showed that the framework does help interpret complex modeling tasks in the laboratory environment.  The framework identified that assumptions were often difficult for students to make explicit, especially when a pre-constructed mathematical model was available.  Further, the interviews provided a striking example of the link between a lack of prior conceptual knowledge and an inability to construct models.  

Because meaningful modeling and experimental design depends on sufficient prior conceptual knowledge, it may be worth revisiting the connection between lecture and lab in the upper-division curriculum. The laboratory experience is commonly viewed as a supplementary experience that aids students' conceptual development, but we are arguing for the community to consider a more complex relationship between conceptual understanding and the experimental process.  There are indications in our data that students' prior conceptual understanding places constraints on the kinds of models they make use of and how they design their experiment.  If scientific practices are to be integrated into science classes at the K-12 and college levels, then we need a clearer understanding of the relationship between conceptual understanding and the learning of scientific practices. 

Future studies will also look at the model construction process, in particular the links between assumptions, limitations, and model revision. Also, we are looking at the relationship between the accuracy and sophistication of students' models of measurement tools and how they connect their own measurements to the key concepts in the lab activity. We hope the framework provides useful insights for those physics education researchers and instructors who are looking for ways to describe laboratory sense-making and for faculty who want to integrate more conceptual and mathematical understanding into experimental activities.

\section{Acknowledgments}

The authors would like to thank the CU Physics Department for input on learning goals and pointing out the importance of modeling measurement tools. The research study was approved by the CU-Boulder Institutional Review Board and informed consent was obtained from all participants.  This work is supported by NSF TUES DUE-1043028 and NSF TUES DUE-1323101. The views expressed in this paper do not necessarily reflect those of the National Science Foundation.

\bibliography{C:/SugarSync/Ben/Bibliographies/Publications-PRST_PER_Modeling_Framework}

\end{document}